\documentclass{optica-article}

\journal{opticajournal} 

\articletype{Research Article}

\usepackage{lineno}

\begin{document}

\title{Modeling of optical scattering from topographic surface measurements of high-quality mirrors}

\author{Tomotada Akutsu\authormark{1,*} and Hiroaki Yamamoto\authormark{2}}

\address{\authormark{1}National Astronomical Observatory of Japan, Mitaka, Tokyo 181-8588, Japan\\
\authormark{2}LIGO Laboratory, California Institute of Technology, Pasadena, California 91125, USA
}

\email{\authormark{*}tomo.akutsu@nao.ac.jp} 


\begin{abstract*}
In this paper, we revisit computational methods to obtain an angular profile of optical scattering from a smooth surface, given a two-dimensional map of topographic height errors of the surface. Quick derivations of some traditional equations and relevant references are organized to shorten the search time. A practical data-processing flow of the methods is discussed. As a case study of this flow, the core mirrors of the KAGRA interferometer are examined, and we obtain a representative scattering profile that is easily applicable to ray-tracing simulations.
\end{abstract*}

\section{Introduction}
Scattered light is a common issue that needs to be addressed for precision measurements using low-loss optical systems.
An extreme example is a gravitational-wave observatory with laser interferometry such as LIGO, Virgo, or KAGRA~\cite{LIGO:2015,Virgo:2015,KAGRA:2020}. The noise due to the scattered light often contaminates the observational data in various modes, and so degrades its sensitivity and reliability~\cite{Abbott:2017,Abac:2024}. Relevant researchers actively continue investigations on the complicated behaviors of the scattered-light noise for the better performance~\cite{Soni:2024,Longo:2024}.

Core optics in all the above-mentioned laser interferometers consist of several high-quality mirrors. They form kilometer-scale Fabry-P\'{e}rot cavities, or \textit{arms}, which are sensitive to the gravitational waves. For each mirror of the arm cavity, the optical loss is very low, 50-100~ppm or even less. The loss is not only due to absorption, but also due to scattering~\cite{Hirose:2020}. Such a tiny amount of scattered light results in critical noise when improperly treated. redFor example, the ``fake'' readout of gravitational-wave amplitude (dimensionless spacetime strain) caused by the scattered-light noise can be expressed as $h\simeq \kappa\,\delta x/l$ in the simplest case, where $l\sim$ a few km is the arm length, $\kappa$ is the fraction of the scattered-light amplitude recombined into the main optical path in the arm cavity, and $\delta x\sim 10^{-8}\,\mathrm{m}$ represents the effective displacement of a secondary scatterer at a few tens of hertz, such as the inner surface of the vacuum enclosure for the arm cavity~\cite{Ottaway:2012}. To suppress the noise-equivalent readout down to $h\sim 10^{-24}$, proper measures must be taken to ensure $\kappa\sim 10^{-12}$ (equivalent to $\kappa^2\sim10^{-24}$ in power ratio or contrast) or lower. The value of $\kappa$ can be estimated through ray-tracing simulations.

To estimate the resultant scattered-light noise, we often start with modeling angular distributions of the scattered light off the relevant high-quality mirrors~\cite{Flanagan:2011,Ottaway:2012,Canuel:2013}. Using the models as optical sources, we can trace the scattered light in the laser interferometer, and design an optical baffle system to mitigate the noise if needed~\cite{Akutsu:2016}. A computational cost for tracing the scattered rays will, however, easily diverge when the models are made too realistic. Therefore, moderately simplified models are more preferable.

This paper summarizes how we obtain such usable models in a traditional manner, and aims to form a rigorous base to the discussion of stray-light mitigation in the future. Section~\ref{Sec2} reviews computational methods to estimate an angular profile of the scattered light off a high-quality mirror from a 2D (two dimensional) measurement of topographic height errors of its surface. These theories are long established~\cite{Stover:1995}. Brief overviews of the relevant mathematics are also included for those curious about the derivations, as we ourselves could not find such things easily.
Section~\ref{Sec3} introduces our practical implementation of those methods, and shows some results when applied to the KAGRA core mirrors as a case study. Here, the main outcome of the case study is an envelope model that represents the scattering profiles of those mirrors, and that will be easily applicable to the analysis including ray-tracging simulations for evaluating the upper limit of the scattered-light noise.
Section~\ref{Sec4} compares the results obtained from the case study with a previous assumption, and also discusses some issues with the current methods. Section~\ref{Sec5} concludes the paper.

\section{Theoretical background}
\label{Sec2}
\subsection{Scattering angular distribution}
For simplicity, let us limit the following discussion on smooth surfaces of low-loss and high-reflective mirrors.
In other words, we assume the power incident on the surface, $\mathcal{P}_\mathrm{i}$, is almost equal to the total scattered-out power $\mathcal{P}_\mathrm{s}$, which includes specular reflection as well as the other scattered light. Note that, in most practical cases, it is difficult or rather meaningless to precisely distinguish the specular and scattered components~\cite{footnote0}.

Under this condition, the scattering probability density function (PDF) per unit solid angle at a certain point of the illuminated surface is
\begin{equation}
\frac{dP}{d\Omega_\mathrm{s}} = \frac{1}{\mathcal{P}_\mathrm{s}}\frac{d\mathcal{P}_\mathrm{s}}{d\Omega_\mathrm{s}}\simeq\frac{1}{\mathcal{P}_\mathrm{i}}\frac{d\mathcal{P}_\mathrm{s}}{d\Omega_\mathrm{s}},
\end{equation}
where $d\Omega_\mathrm{s}$ indicates a fraction of solid angle around a scattered angle in concern, which can be expressed by a combination of a polar angle $\theta_\mathrm{s}$ and an azimuthal angle $\phi_\mathrm{s}$, while $d\mathcal{P}_\mathrm{s}$ is the light power scattered within $d\Omega_\mathrm{s}$. The angle of incidence on the illuminated surface can be also expressed by a similar combination $(\theta_\mathrm{i}, \phi_\mathrm{i})$, but proper choices of the coordinates can set $\phi_\mathrm{i}=0$ without loss of generality in most cases. Note that $\int dP/d\Omega_\mathrm{s}\, d\Omega_\mathrm{s}=1$ due to this definition or the energy conservation, where the integration is taken over the whole sphere around the concerned point on the illuminated surface. By the way, the traditional BSDF (bi-directional scatter distribution function) corresponds to $(d\mathcal{P}_\mathrm{s}/d\Omega_\mathrm{s})/(\mathcal{P}_\mathrm{i}\cos\theta_\mathrm{s})$, which is approximately equal to the scattering PDF when $|\theta_\mathrm{s}|\ll 1$.

The traditional Rayleigh-Rice theory relates the topographic height errors of the illuminated surface to the scattering distribution as (see Ref.~\cite{Stover:1995})
\begin{align}
    \frac{1}{\mathcal{P}_\mathrm{i}}\frac{d\mathcal{P}_\mathrm{s}}{d\Omega_\mathrm{s}}=\frac{16\pi^2}{\lambda^4}\cos^2\theta_\mathrm{s}\cos\theta_\mathrm{i}\,Q\,S_2(f_x,f_y),\label{eq:RRtheory}
\end{align}
where $\lambda$ is a wavelength of the light, $Q$ is a function of $\theta_\mathrm{s}$,$\phi_\mathrm{s}$, and $\theta_\mathrm{i}$ to describe a distribution of the mean light power, and $S_2$ is a two-sided 2D (two dimensional) power spectral density (PSD) for a map of the topographic height errors of the surface. $S_2$ is also a function of the angles, because the spatial frequencies $f_x$ and $f_y$ are related with the angular parameters:
$f_x = (\sin\theta_\mathrm{s}\cos\phi_\mathrm{s}-\sin\theta_\mathrm{i})/\lambda$, and 
$f_y = (\sin\theta_\mathrm{s}\sin\phi_\mathrm{s})/\lambda$, respectively. Further details on $S_2$ are provided in Section~\ref{Sec.S2}.

For the later discussion, here we derive the more reduced forms with some assumptions. Let $\theta_\mathrm{i}\simeq 0$. 
Recall the assumption $\mathcal{P}_\mathrm{s}\simeq \mathcal{P}_\mathrm{i}$. Assume that most of the scattered light power is distributed around the direction of the specular reflection, allowing us to consider only the case of $|\theta_\mathrm{s}|\ll 1$. These assumptions also imply $Q\sim \mathcal{P}_\mathrm{s}/\mathcal{P}_\mathrm{i}\sim 1$, as $Q$ is a generalized reflectivity in the context of this paper.
Then, the scattering PDF becomes $dP/d\Omega_\mathrm{s}\simeq 16\pi^2 S_2(f_x,f_y)/\lambda^4$, where $f_x\simeq \theta_\mathrm{s}\cos\phi_\mathrm{s}/\lambda$ and $f_y\simeq \theta_\mathrm{s}\sin\phi_\mathrm{s}/\lambda$.
Furthermore, if $S_2$ is circularly symmetric, they reduce to
\begin{equation}
\frac{dP}{d\Omega_\mathrm{s}}(\theta_\mathrm{s})\simeq\frac{16\pi^2}{\lambda^4}\,S_2(f),\label{eq:dP/dW}
\end{equation}
where
\begin{equation}
f\equiv(f_x^2+f_y^2)^{1/2}\simeq\theta_\mathrm{s}/\lambda\label{eq:f2theta}.
\end{equation}

\subsection{Basics of 2D transforms}
\label{Sec.S2}
\subsubsection{General descriptions}
When the surface topographic height errors are measured and then mapped as $z(x,y)$, the (size-limited) 2D Fourier transform of this map is
\begin{equation}
Z_L(f_x, f_y)\equiv \int_{-L/2}^{L/2}\int_{-L/2}^{L/2} z(x,y)\, e^{-i 2\pi (f_x x + f_y y)}\,dx\,dy\label{eq:Z_L}
\end{equation}
where $L$ is a (tentatively limited) size of the map, and $-\infty<f_x,f_y<\infty$ if this double integral converges.
Note that $Z_L(-f_x,-f_y)=Z_L^*(f_x,f_y)$, and also $Z_L(-f_x,f_y)=Z_L^*(f_x,-f_y)$; however, there is not always this kind of symmetry between $Z_L(f_x,f_y)$ and $Z_L(-f_x,f_y)$.

The two-sided 2D PSD associated with the map $z(x,y)$ is
\begin{equation}
S_2(f_x, f_y) = \lim_{L\rightarrow\infty}\left\langle\frac{1}{L^2}\,|Z_L(f_x,f_y)|^2\right\rangle\label{eq:S2},
\end{equation}
where $-\infty<f_x,f_y<\infty$ if the limit exists, and $\langle\cdot\rangle$ means the ensemble average. Note that $S_2(-f_x,-f_y)=S_2(f_x,f_y)$, and also $S_2(-f_x, f_y)=S_2(f_x, -f_y)$; however, there is not always this kind of symmetry between $S_2(f_x,f_y)$ and $S_2(-f_x,f_y)$.

\subsubsection{Abel transforms}
While the following formulas are widely used, some of the mathematical derivation cannot be easily found. For the future usefulness, here we briefly outline the derivations, or provide links to them.

For the simpler expression, the 2D PSD can be converted or \textit{compressed} into a 1D PSD by allowing some information loss. The following method is widely used~\cite{Church:1989,Stover:1995}:
\begin{equation}
S_1(f_x)=2\int_{-\infty}^\infty S_2(f_x,f_y)\,df_y\label{eq:S1}
\end{equation}
where $S_1(f_x)$ is defined as a \textit{one-sided} 1D PSD for $f_x \in (0,+\infty)$, so the factor 2 is added in the right-hand side~\cite{footnote1}.
When the topographic map $z(x,y)$ is isotropic, its two-sided 2D PSD is circularly symmetric~\cite{footnote1.5}, and Eq.(\ref{eq:S1}) can take a particular form, the (forward) Abel transform~\cite{Church:1989,Stover:1995,Bracewell:2003}:
\begin{equation}
    S_1(f_x)=4\int^\infty_{f_x}\frac{fS_2(f)}{\sqrt{f^2-f_x^2}}\,df,\label{eq:S1_abel}
\end{equation}
where $f\equiv (f_x^2+f_y^2)^{1/2}$ for $f_x \in (0,\infty)$. The derivation is simple; divide the integration range in the right-hand side of Eq.~(\ref{eq:S1}) into $\int_{-\infty}^0+\int_0^\infty$ for $f_y=\pm (f^2-f_x^2)^{1/2}$, convert respectively via $df_y=\pm (f^2-f_x^2)^{-1/2}f\,df$, and use the circular symmetric feature $S_2(f_x,f_y)=S_2(f)$.
If necessary, the inverse Abel transform is given~\cite{Church:1989,Stover:1995}:
\begin{equation}
S_2(f) = -\frac{1}{2\pi}\int^\infty_f\frac{1}{\sqrt{f_x^2-f^2}}\frac{dS_1(f_x)}{df_x}\,df_x\label{eq:S1toS2}
\end{equation}
for $f \in (0,\infty)$. The derivation is found in~\cite{Bracewell:2000}.

In the early days, only 1D profiling tools were sometimes available, and so $S_2$ was reconstructed from $S_1$ by assuming the circular symmetry of $S_2$. Nowadays, 2D profiling tools are more commonly available.

\subsection{Spectral models}
In the following, two widely-used spectral models are introduced: the inverse power law model and the ABC model. Mathematically, the former can be understood as a certain approximation of the latter.

\subsubsection{Inverse power law model}
The inverse power law model is simple~\cite{Church:1988}
\begin{equation}
S_1(f_x) = K f_x^{-n}\quad (f_x>0),\label{eq:S1_powerlaw}
\end{equation}
with $1<n<3$ and $K>0$,
and its inverse Abel transform is
\begin{equation}
S_2(f) = \frac{\Gamma((n+1)/2)}{2\sqrt{\pi}\,\Gamma(n/2)}\,K f^{-(n+1)}\quad (f>0),\label{eq:S2_powerlaw}
\end{equation}
where $\Gamma(\cdot)$ is the gamma function (see Chap.~5.2 of~\cite{NIST:DLMF}).

The derivation of Eq.(\ref{eq:S2_powerlaw}) from Eq.(\ref{eq:S1_powerlaw}) is as follows.
Substitute $dS_1/df_x=-nKf_x^{-(n+1)}$ to Eq.(\ref{eq:S1toS2}), and convert the integration variable with $\eta\equiv (f/f_x)^2$ via $df_x = -\frac{1}{2}f\eta^{-3/2} d\eta$, where the integration range $f_x: f\rightarrow \infty$ changes to $\eta: 1\rightarrow 0$. Then $S_2(f) = \frac{K}{4\pi}nf^{-(n+1)}\mathcal{B}(\frac{n+1}{2},\frac{1}{2})$ is obtained, where $\mathcal{B}(x,y)=\int_0^1 t^{x-1} (1-t)^{y-1}dt$ is the beta function, and a relation $\mathcal{B}(x,y)=\Gamma(x)\Gamma(y)/\Gamma(x+y)$ is known (see Chap.~5.12 of~\cite{NIST:DLMF}). Substituting $\Gamma(\frac{1}{2})=\sqrt{\pi}$ and $\Gamma(x+1) = x\,\Gamma(x)$ for $x>0$ yields Eq.(\ref{eq:S2_powerlaw}).

\subsubsection{ABC model}
The ABC model, or K-correlation model, is also traditional  and widely used for modeling the scattering profiles~\cite{Church:1989,Stover:1995,HYamamoto:2011}. The core of the idea is to regard the surface topographic height errors as a statistical quantity having an autocorrelation of a family of the modified Bessel functions, which are often expressed as $K(\cdot)$; see the early discussions~\cite{Noll:1979,Noll:1980,Church:1980,Noll:1982,Church:1982,Church:1986}. In statistical terms, this model corresponds to the Pearson type VII distribution~\cite{footnote2}, or the non-standardized Student's $t$ distribution.

The general form is, with the parameters $A, B, C>0$,
\begin{equation}
S_1(f_x)=\frac{A}{[1+(Bf_x)^2]^{(C/2)}}\quad (f_x>0),\label{eq:S1_ABC}
\end{equation}
and~\cite{footnote3} its inverse Abel transform is
\begin{equation}
S_2(f)=\frac{A'}{[1+(Bf)^2]^{(C+1)/2}}\quad (f>0),\label{eq:S2_ABC}
\end{equation}
where $A'\equiv\frac{\Gamma ((C+1)/2)}{2\sqrt{\pi}\,\Gamma(C/2)}AB$.

The derivation of Eq.(\ref{eq:S2_ABC}) from Eq.(\ref{eq:S1_ABC}) follows the same procedure as before, up to the point where the integration variable is converted to $\eta$. Next, replace $\eta$ further with $t\equiv 1-\eta$, and then $S_2(f)=I_0\frac{AB^2C}{2\pi}\frac{f}{2}u^{(\frac{C}{2}+1)}$, where
$I_0\equiv \int_0^1 t^{-\frac{1}{2}}\cdot (1-t)^{\frac{C-1}{2}}\cdot (1 -tu)^{-(\frac{C}{2}+1)}\, dt$ and $u\equiv [1+(Bf)^2]^{-1}$, is obtained.
With the hypergeometric function $F(\alpha,\beta;\gamma;z)$, see (15.6.1) with (15.2.1) of~\cite{NIST:DLMF}, the integral reduces to $I_0 = \frac{2}{C}\frac{\sqrt{\pi}\,\Gamma((C+1)/2)}{\Gamma(C/2)}F(\frac{C}{2}+1,\frac{1}{2};\frac{C}{2}+1; u)$. By referring to (15.4.6) of~\cite{NIST:DLMF}, $F(\alpha,\beta;\alpha;z) = (1-z)^{-\beta}$, and so $I_0= \frac{2}{C}\frac{\sqrt{\pi}\,\Gamma((C+1)/2)}{\Gamma(C/2)}(1-u)^{-\frac{1}{2}}$, which yields Eq.~(\ref{eq:S2_ABC}).

Note that this model converges to the inverse power law model in Eq.~(\ref{eq:S1_powerlaw}) or (\ref{eq:S2_powerlaw}) with $K\simeq A/ B^C$ and $n\simeq C$, when $(Bf_x)^2\gg 1$.

\section{Data analysis and results}
\label{Sec3}
In this section, a data processing flow and the results are discussed. As already mentioned, our objective is to estimate an optical scattering profile of a mirror surface from its measured topographic height map, and to make this profile easily available for various ray-tracing simulations in the future.

\subsection{Data processing}
\begin{figure}[tb!]
\centering\includegraphics[width=4.5cm]{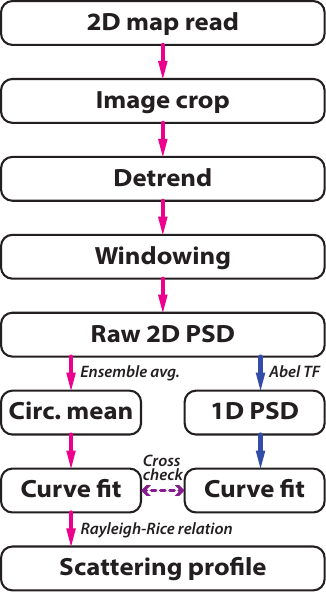}
\caption{Summary of the data processing flow.}
\label{fig:flow}
\end{figure}

Fig.~\ref{fig:flow}
summarizes our data processing flow.
The flow is implemented into a \verb|python| script.
As discussed later, this time the read data are the measured 2D maps of the KAGRA mirrors, which had been characterized in~\cite{Hirose:2020}.
The map data are saved in Zygo's \verb|.dat| format, and so the \verb|prysm| library~\cite{Dube2019} is utilized to read them.

For pre-processing, each map image is cropped to be a 140-mm diameter circle centered at the image's center. Then, the cropped map is detrended with the \verb|poppy| library, an open-source optical propagation Python package originally developed for the James Webb Space Telescope (JWST) project~\cite{2016ascl.soft02018P,2012SPIE.8442E..3DP}; specifically, components corresponding to the typical low-order Zernike polynomials ($Z_1$, $Z_2$, $Z_3$, and $Z_4$ in Table~I of~\cite{Noll:1976}) are evaluated, and then subtracted.

Before performing 2D FFT (fast Fourier transform) for the pre-processed map, a 2D Hann window is applied. In the normalized form, it is written as
\begin{equation}
w(x,y)=\tfrac{1}{2}-\tfrac{1}{2}\cos\left[2\pi (t-\tfrac{1}{2})\right]
\end{equation}
where $x=\frac{1}{2}+t\cos\phi$ and $y=\frac{1}{2}+t\sin\phi$ for $0\leq t\leq\frac{1}{2}$ and $0\leq\phi<2\pi$, while $w(x,y)=0$ for the other $t$. It takes the maximum 1 at $(x,y)=(\frac{1}{2},\frac{1}{2})$. Note that the domain of the window function should be scaled to the map size in the real coding. The corresponding window is introduced with the \verb|scikit-image| library\cite{scikit-image}, and then multiplied with the map.

After performing 2D FFT for the windowed map, a raw 2D PSD is calculated with
\begin{equation}
s_2=\frac{1}{c_wL^2}|Z_L^\diamond|^2,\label{eq:raw_2D_PSD}
\end{equation}
where $Z_L^\diamond$ is the practical discrete version of $Z_L$ in Eq.~(\ref{eq:Z_L}). The window correction factor $c_w$ is calculated as
$\int_0^1\int_0^1w^2(x,y)\,dx\,dy$, which corresponds to $2\pi \int_0^{1/2} t\, w^2(t)\,dt$, or $c_w\sim 0.135$.
Note that $s_2$ is not ensemble averaged yet, so this spectrum has a large estimation error.

One way to obtain $S_2^\diamond$, which is the practical discrete version of Eq.~(\ref{eq:S2}), or the ensemble average of $s_2$, is to calculate a circular mean of $s_2$ under the assumption that the original mirror map $z(x,y)$ is isotropic. This implies the corresponding two-sided 2D PSD should be circularly symmetric. The circular mean can be formally written as
\begin{equation}
S_2^\diamond(f)=\frac{1}{N}\sum_{k=0}^{N-1} s_2(f_x,f_y),
\label{eq:S2_diamond}
\end{equation}
where $f_x=f\cos(2\pi k/N)$ and $f_y=f\sin(2\pi k/N)$ for $f>0$, and $N$ is a positive integer for averaging. Note that $S_2^\diamond$ is still a \textit{two-sided} 2D PSD estimate, although $f>0$ is put in the calculation above; in other words, $S_2^\diamond(f_x,f_y) = S_2^\diamond(f)$.

From $S_2^\diamond$, relevant parameters in Eq.~(\ref{eq:S2_powerlaw}) or (\ref{eq:S2_ABC}) are estimated by curve fitting. In this way, the two-sided 2D PSD for the measured mirror map is modeled by a simple function with a few parameters. Using these parameters, the scattering PDF estimate can be obtained through Eq.~(\ref{eq:dP/dW}).

In Fig.~\ref{fig:flow}, there is another branch from the ``Raw 2D PSD'' node. This path compresses the 2D PSD to a 1D PSD using Eq.~(\ref{eq:S1}). Under the assumption that the 2D PSD is circularly symmetric, this corresponds to the Abel transform shown in Eq.~(\ref{eq:S1_abel}). Furthermore, the compression automatically results in an averaging effect at the same time. This way, $S_1^\diamond(f_x)$, which is a practical discrete version of Eq.~(\ref{eq:S1}), is obtained from $s_2(f_x,f_y)$. Note that $S_1^\diamond(f_y)$ can also be obtained by integrating over $f_x$. In our case, the average $(S_1^\diamond(f_x)+S_1^\diamond(f_y))/2$ was adopted for the one-sided 1D PSD estimate. From $S_1^\diamond$ data, relevant parameters in Eq.~(\ref{eq:S1_powerlaw}) or (\ref{eq:S1_ABC}) are estimated by curve fitting. These estimated parameters can be compared with the relevant parameters estimated from $S_2^\diamond$ described above, for consistency check.

\subsection{Application and results}
\label{Sec.Appl.Res}
This subsection presents the results obtained by applying the data processing flow in Fig.~\ref{fig:flow} to the measured 2D topographic surface maps of the KAGRA core mirrors as a case study.

\begin{figure}[tb!]
\centering\includegraphics[width=14cm]{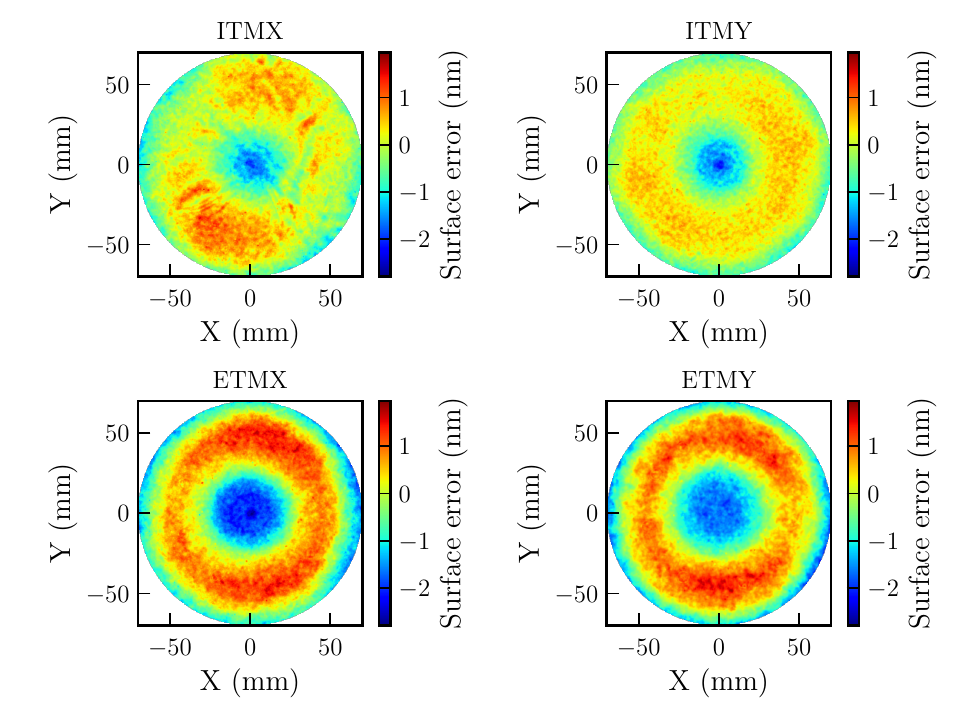}
\caption{Detrended surface error maps of the KAGRA core mirrors~\cite{Hirose:2020}.}
\label{fig:detrend}
\end{figure}

Fig.~\ref{fig:detrend} shows the outcome of the ``Detrend'' node.
The KAGRA interferometer has four core mirrors, named ITMX, ITMY, ETMX, and ETMY, respectively~\cite{footnote4}; the details for their measurement and characterization are found in~\cite{Hirose:2020}, where a Fizeau interferometer at Caltech (a customized Zygo Verifire) was used to measure the surface figure error of the mirrors with respect to a reference sphere using the phase-shifting interferometry (PSI) method. 

In fact, the method to obtain Fig.~\ref{fig:detrend} is conceptually the same as the one to obtain the relevant figure in~\cite{Hirose:2020}. In Fig.~\ref{fig:detrend}, each map image is cropped to a 140-mm diameter centered at the mirror center, and the corresponding image data is 351 $\times$ 351 pixels. In addition, as mentioned above, the first four Zernike terms are subtracted. Among them, the first three terms arise from the mis-positioning of the mirror under test in the Fizeau interferometer, while the fourth term corresponds to the residual curvature component relative to the reference sphere. By removing these components, the intrinsic figure error of the mirror surface (i.e. the sum of the remaining higher-order Zernike terms; see also Ref.~\cite{Bond:2017}), becomes apparent, as shown in Fig.~\ref{fig:detrend}. Note that the removed components affect the PSD estimate only in the lowest spatial frequency range, and did not significantly alter the fit results discussed later.

\begin{figure}[tb]
\centering\includegraphics[width=14cm]{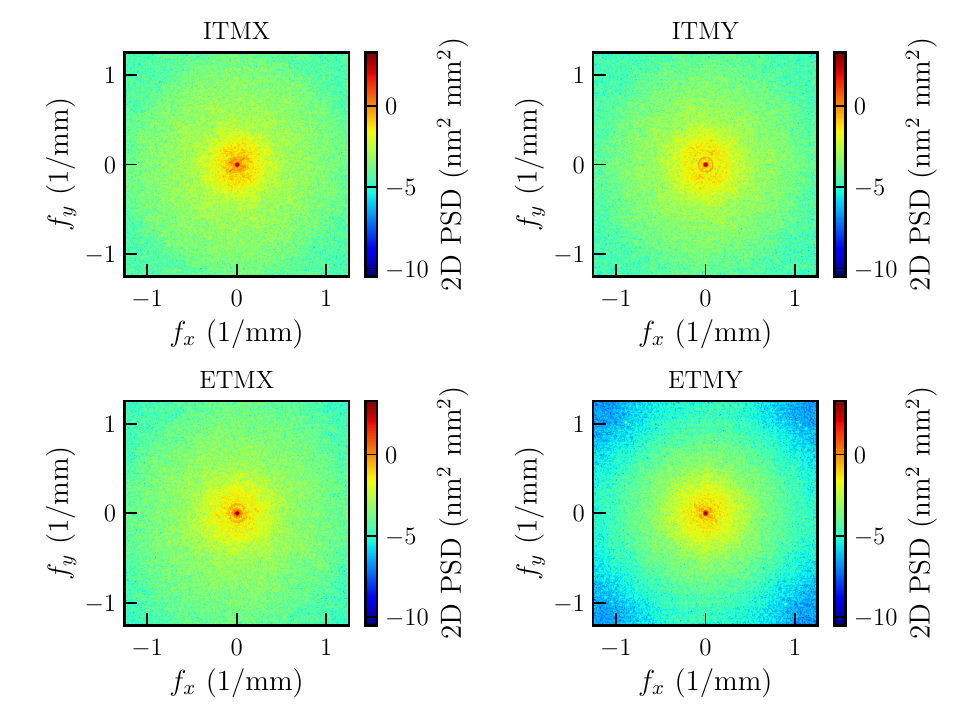}
\caption{Raw two-sided 2D power spectral density (PSD) maps of the KAGRA core mirrors. Each color bar represents the exponent of 10.}
\label{fig:2DPSD_map}
\end{figure}

Fig.~\ref{fig:2DPSD_map}
shows the outcome of the ``Raw 2D PSD'' node, or $s_2$ in Eq.~(\ref{eq:raw_2D_PSD}).
For each mirror, the corresponding $s_2$ is plotted with respect to the spatial frequencies $f_x$ and $f_y$. The color bar represents the 2D PSDs in terms of the exponent of 10, and the color scale is common for every subplot. The circular symmetry of each $s_2$ is apparent.

\begin{figure}[tb]
\centering\includegraphics[width=14cm]{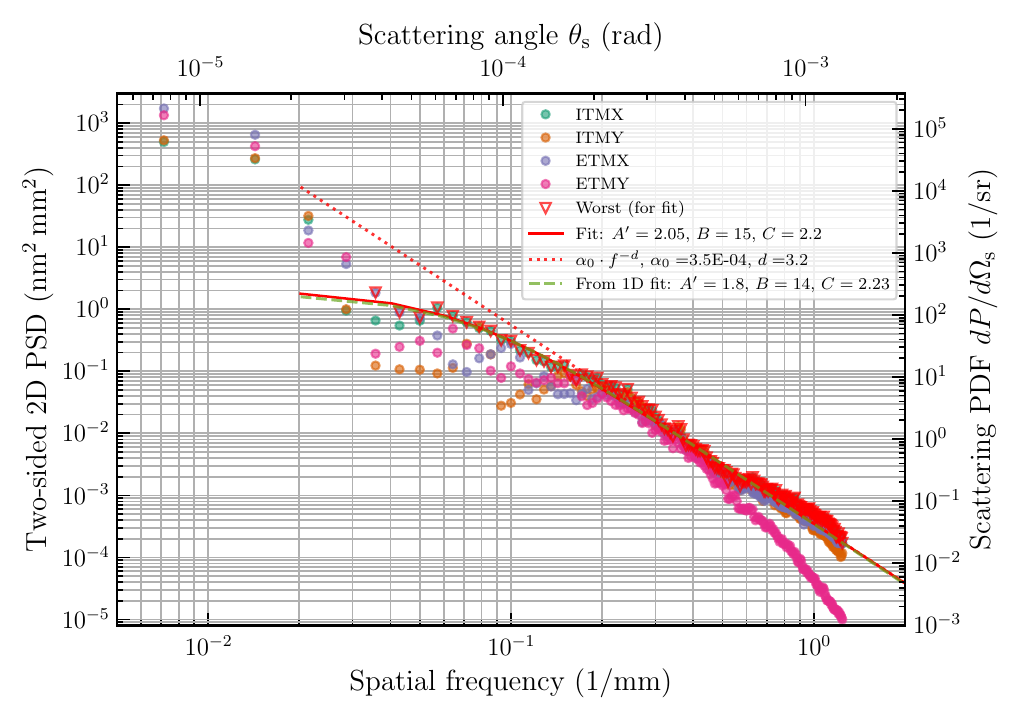}
\caption{Equivalent \textit{two-sided} 2D power spectral density (PSD) and the corresponding scattering distribution function.}
\label{fig:2DPSD_BRDF}
\end{figure}

Outcomes of the ``Circ. mean'' node, or $S_2^\diamond$ in Eq.(\ref{eq:S2_diamond}) for the four mirrors are plotted in Fig.~\ref{fig:2DPSD_BRDF}
with filled circles. Again note that each 2D PSD is in two-sided form. Then, the worst values among the four data sets at each spatial frequency are marked with upside-down triangles in the range above $0.03\,\mathrm{mm^{-1}}$.

To perform curve fit, the lower limit of the spatial frequency is set. This threshold is derived as follows. For the KAGRA's core mirrors, scattered light ($\lambda=1064\,\mathrm{nm}$) to be considered is that spreads more than $110\,\mathrm{mm}/3\,\mathrm{km} \sim 37\,\mathrm{\mu rad}$ in the polar angle~\cite{Akutsu:2016}, where the length of the main Fabry-P\'{e}rot arm cavity is 3~km, while each core mirror at the both ends of the cavity is 110~mm in radius. This corresponds to $\sim 0.035\,\mathrm{mm^{-1}}$ according to Eq.~(\ref{eq:f2theta}). We use 2D PSD values above $\sim 0.03\,\mathrm{mm^{-1}}$ for the curve fit.

The fit curve is drawn with a solid curve in Fig.~\ref{fig:2DPSD_BRDF}.
The fit data are indicated with upside-down triangles, which are the worst-case values among the four mirrors, as already mentioned. This is because it is enough to discuss on the worst case or the upper limit of the noise due to the scattered light for our purpose. Recall that our objective is to obtain an estimate of scattering profile that is available for various ray-tracing simulations in the future, but not to characterize each mirror's quality. The real world is quite complicated for this kind of simulation, so certain simplifications and assumptions are unavoidable to obtain usable results efficiently. For example, one of the apparent differences is that rays are described in geometrical optics while the laser light in use is well described in wave optics. Then, the results include various uncertainties. Practically, we further include a safety factor of 10 to 100 (depending on the situation) into the baffle design in the end. Note that, despite these treatments, currently operating gravitational-wave detectors such as LIGO, Virgo, and KAGRA still suffer from the stray-light noise through unforeseen coupling paths.

In this way, outcomes of the left ``Curve fit'' node, or fit parameters can be obtained as $A'=2.049\pm 0.141$, $B=15.00\pm 1.22$, and $C=2.198\pm 0.132$ in Eq.~(\ref{eq:S2_ABC}). Note that $A'$ and $1/B$ have the same dimension as the 2D PSD ($\mathrm{nm^2\, mm^2}$) and the spatial frequency ($\mathrm{mm^{-1}}$), respectively, while $C$ is dimensionless.
In the actual coding, the curve fit is weighted by the spatial frequency to reduce the estimation errors.

The dotted line in Fig.~\ref{fig:2DPSD_BRDF} is calculated from the fit curve described above as a power-law limit; $(Bf)^2\gg 1$ in Eq.~(\ref{eq:S2_ABC}). This corresponds to Eq.~(\ref{eq:S2_powerlaw}). For a more conservative upper limit of the scattering profile, the inverse power law estimation may also be useful in certain situations.

\begin{figure}[tb]
\centering\includegraphics[width=14cm]{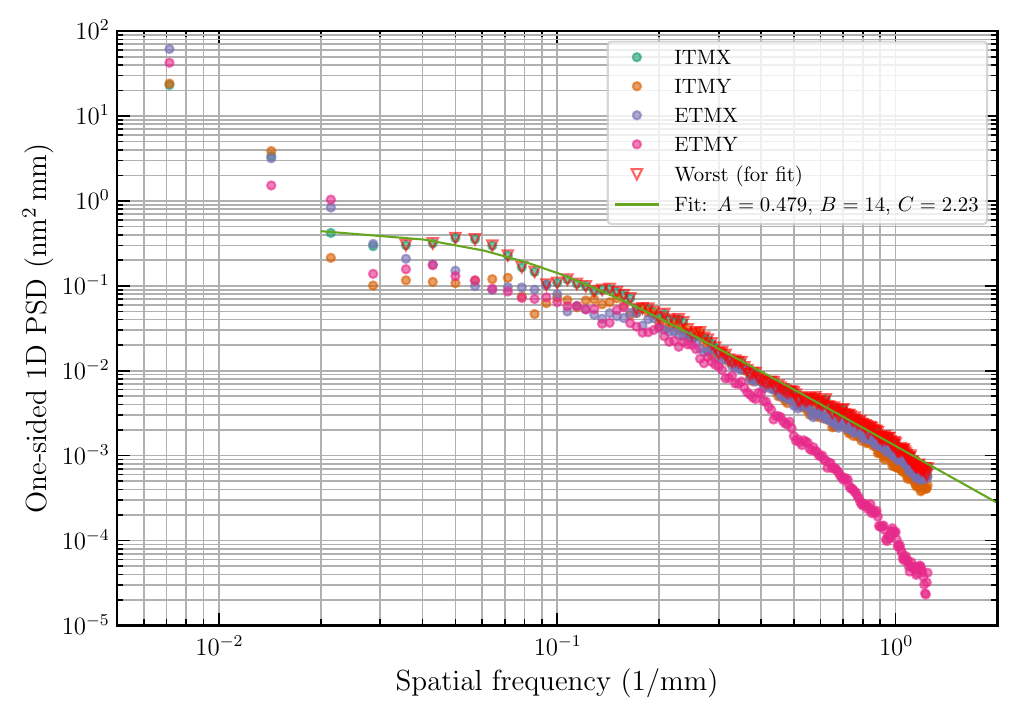}
\caption{One-sided 1D power spectral density (PSD).}
\label{fig:1DPSD}
\end{figure}

Outcomes of the ``1D PSD'' node, or $(S_1^\diamond(f_x)+S_1^\diamond(f_y))/2$, are plotted in
Fig.~\ref{fig:1DPSD}
for ITMX, ITMY, ETMX, and ETMY, respectively. Note that the vertical axis is the \textit{one-sided} 1D PSD. In the same manner as the 2D PSD case, the worst-case values above the threshold spatial frequency are marked with upside-down triangles. The curve fit is performed for these data with Eq.~(\ref{eq:S1_ABC}).

In the same manner as the 2D PSD case, outcomes of the right ``Curve fit'' node are obtained as $A=0.479\pm 0.031$, $B=14.04\pm 0.69$, and $C=2.234\pm 0.043$ for Eq.~(\ref{eq:S1_ABC}). The curve fit is weighted by the spatial frequency to reduce the estimation errors.
From the fit result, the relevant $S_2(f)$ in Eq.~(\ref{eq:S2_ABC}) can be also reconstructed. For cross check, this is drawn in Fig.~\ref{fig:2DPSD_BRDF} by a dashed line. Compared with the solid line, the agreement is reasonable. This fact supports the assumption that the original two-sided 2D PSD is of circular symmetry (also supported by Fig.~\ref{fig:2DPSD_map}), and suggests the scattering profile is nearly independent of the azimuthal angles.

As an outcome of ``Scattering profile'' node, the scattering PDF $dP/d\Omega_\mathrm{s}$ can be calculated. To show this, the right vertical axis in Fig.~\ref{fig:2DPSD_BRDF} is converted to this scale according to Eq.~(\ref{eq:dP/dW}). Also, the scattering angle (polar angle $\theta_\mathrm{s}$) that corresponds to the spatial frequency is shown on the upper horizontal axis. Note that the scattering PDF here is of the isotropic surfaces, so should be invariant under the azimuthal angle $\phi_\mathrm{s}$ changes. Consequently, the obtained model for the optical scattering profile is expressed using the ABC model as
\begin{equation}
\frac{dP}{d\Omega_\mathrm{s}}\simeq 2.5\times10^2\cdot\left[1+\left(\frac{\theta_\mathrm{s}}{7.1\times 10^{-5}\,\mathrm{rad}}\right)^2\right]^{-1.6}\quad\mathrm{[1/sr].}~\label{eq:resultant1}
\end{equation}
Alternatively, in terms of the power law model, it is given by
\begin{equation}
    \frac{dP}{d\Omega_\mathrm{s}}\simeq 1.3\times 10^{-11}/\theta_\mathrm{s}^{3.2}\quad\mathrm{[1/sr]}
    \label{eq:resultant2}.
\end{equation}

\section{Discussion}\label{Sec4}
\subsection{Comparison with the previous model}
Eqs.~(\ref{eq:resultant1}) and (\ref{eq:resultant2}) provide the first model-based estimation of the optical scattering profile $dP/d\Omega_\mathrm{s}$ for the KAGRA core mirrors. An alternative approach would be direct measurement using an optical goniometer, but such measurements have not been conducted. In this subsection, we compare these profiles with the one previously assumed for designing the optical baffle system of the KAGRA arm: $dP/d\Omega_\mathrm{s}\simeq 2\times 10^{-6}/\theta_\mathrm{s}^2$ [1/sr]. At the time this assumption was made, the KAGRA core mirrors had not yet been manufactured, and no topographic data of their surfaces were available.

The earlier estimate of scattered-light noise in the KAGRA arm was based on past studies of LIGO's silica mirrors~\cite{Flanagan:2011,Weiss:1998}. It was assumed that $dP/d\Omega_\mathrm{s}\propto\theta_\mathrm{s}^{-2}$ would hold for the KAGRA mirrors, given that a similar manufacturing quality was expected despite the different substrate material, sapphire~\cite{Hirose:2020}. Using this profile, ray-tracing simulations were performed within the arm's vacuum enclosure to estimate the fraction of scattered light recombining into the main optical path. The optical baffle shapes and locations were then designed accordingly. The coefficient $2\times 10^{-6}$ was later derived from test-polished samples, enabling the scaling of the scattered-light noise evaluation.

The updated profiles refine these estimates. The previous profile overestimated the scattered light at angles $\theta_\mathrm{s}> 48\,\mathrm{\mu rad}$ compared with Eq.~(\ref{eq:resultant2}), while only components with $\theta_\mathrm{s} > 37\,\mathrm{\mu rad}$ are relevant (see Section~\ref{Sec.Appl.Res}). In the range $37\,\mathrm{\mu rad} < \theta_\mathrm{s} < 48\,\mathrm{\mu rad}$, it also overestimated the scattered light relative to Eq.~(\ref{eq:resultant1}).
This suggests that the scattered-light noise in the KAGRA arm is less significant than previously estimated, requiring no immediate changes to the baffle design. If a major upgrade to the optical system within the arm, such as a modification of the photon calibrator~\cite{Inoue:2023}, is required in the future, the noise re-evaluation should be conducted based on the updated profiles.

\subsection{Issues and limitations}
We recognize some issues regarding the current methods described above, and discuss them in this subsection. Fortunately, LIGO and Virgo succeeded in showing the traditional approach is sufficient to start gravitational-wave astronomy. For further improved observations, however, these issues may need to be revisited.

Currently, every core mirror in the interferometers has a multilayer coating on its each surface, while the analysis above implicitly assumes an uncoated mirror. The multilayer coating controls the reflectivity and/or transmissivity of the mirror for incident laser light at specific wavelengths and angles of incidence. Due to the multilayer, it is not obvious whether the actual scattering angular profile would follow the relation in Eq.~(\ref{eq:RRtheory}); only a limited number of studies have been conducted on this topic so far~\cite{Zeidler:2017}.

As mentioned above, Eq.~(\ref{eq:dP/dW}) for $|\theta_\mathrm{i}|\simeq 0$ yields the scattering profile $dP/d\Omega_\mathrm{s}$ shown in Fig.~\ref{fig:2DPSD_BRDF}. In the ray-tracing simulations, however, the mirror itself is sometimes illuminated again by the scattered-light rays at arbitrary angles of incidence due to multiple reflections in the optical system. Fortunately, a theory in~\cite{Flanagan:2011}, the reciprocity relation, mostly resolves this concern; one can compute how much of such scattered light will recombine into the main optical path. The remaining concern is the other scattered light that will not be recombined here. For example, some may pass through the mirror substrate and exit from the other side. Probably, the number of them would be small and ignorable, but we have no definite idea of their treatment so far.

The following is not directly related to the validity of the methods, but it is worth noting for future applications. The case study above used the figure-error maps of the KAGRA core mirrors as input data. Due to the limited resolution of the maps, the resultant PSD estimates obtained were less than $\sim 1\,\mathrm{mm^{-1}}$, as shown in Figs.~\ref{fig:2DPSD_BRDF} and \ref{fig:1DPSD}. This corresponds to $\theta_\mathrm{s}$ less than $\sim 1\,\mathrm{mrad}$ for $dP/d\Omega_\mathrm{s}$. To cover the full region of the arm cavity, however, we ideally need data up to $\sim 0.1\,\mathrm{rad}$ (i.e. around a few degrees), or $\sim 100\,\mathrm{mm^{-1}}$ in terms of the spatial frequency, except for the wider angular region that would be dominated by Lambert-like diffusive scattering due to point defects in the coating of the mirror. Using a microscope, one can measure the surface roughness of the mirror substrate before being coated, or the components in the higher spatial frequency $\sim 100\,\mathrm{mm^{-1}}$~\cite{Hirose:2020}. However, it remains unclear what is indicated with a microscopic measurement of the mirror surface after multilayer coating.

\section{Conclusion}
\label{Sec5}
We reviewed theories of traditional methods for estimating the angular profile of optical scattering off a high-quality mirror from topographic height errors on the mirror surface. We implemented the methods into a script. A case study was performed using the script for core mirrors of KAGRA, a gravitational-wave observatory in Japan, and we obtained a representative model of the scattering profiles. Despite some issues with the current methods, the resultant model will be useful for future evaluations of scattered-light noise at KAGRA. The same method will be applicable for future interferometers.


\begin{backmatter}
\bmsection{Funding}
(place holder. According to the style guide, this section will be automatically generated upon submission to the system and the text here will be ignored.) 

\bmsection{Acknowledgments}
This work was supported by Ministry of Education, Culture, Sports, Science and Technology (MEXT), Japan Society for the Promotion of Science (JSPS) Leading-edge Research Infrastructure Program, JSPS Grant-in-Aid for Specially Promoted Research 26000005, and the joint research program of the Institute for Cosmic Ray Research, University of Tokyo. KAGRA is supported by MEXT and JSPS in Japan; National Research Foundation (NRF) and Ministry of Science and ICT (MSIT) in Korea; Academia Sinica (AS) and National Science and Technology Council (NSTC) in Taiwan. We acknowledge the cooperative support of the LIGO Laboratory of the National Science Foundation (NSF), Caltech, and the use of its facilities for characterizing the KAGRA core optics. We acknowledge GariLynn Billingsley, Liyuan Zhang, and Eiichi Hirose for the measurement and data reduction of the mirror maps; Yoichi Aso, Kentaro Somiya, and Haoyu Wang for suggesting us useful tips to implement the data processing; Matteo Leonardi, Marc Eisenmann, and Yuta Michimura for helping the mirror data salvage.

\bmsection{Disclosures}
The authors declare no conflicts of interest.

\bmsection{Data availability}
Data underlying the results presented in this paper are not publicly available at this time but may be obtained from the authors upon reasonable request.

\end{backmatter}

\bibliography{MirrorProfile2ScatteringProfile}

\begin{thebibliography}{10}
\bibitem{footnote0}
Multiple factors contribute to this difficulty, including the limited angular
  resolution of an optical goniometer, where a photodetector is illuminated by
  specularly reflected light from a surface under test. Due to the finite
  light-sensitive area of the photodetector, the measurement is contaminated by
  scattered light surrounding the specular reflection, and making it difficult
  to isolate the light component reflected at exactly the same angle of the
  incident beam.
\bibitem{footnote1.5}
In general, the 2D PSD $S_2$ of a map $z(x,y)$ is obtained by applying the
  Fourier transform to the autocorrelation of $z(x,y)$. For isotropic maps, by
  definition, the autocorrelation depends solely on the distance between any
  two points, given by $\delta=((x_1-x_2)^2+(y_1-y_2)^2)^{1/2}$, and not on the
  directional angle between them. Consequently, the corresponding $S_2$ depends
  solely on a parameter with units of $\delta^{-1}$, commonly known as spatial
  frequency. As a result, this $S_2$ exhibits circular symmetry when
  represented in the 2D spatial frequency domain.
\end{thebibliography}

\end{document}